\documentclass[a4paper,10pt,english]{article}
\usepackage[margin=1in]{geometry}
\usepackage[utf8]{inputenc}
\usepackage{enumitem}
\PassOptionsToPackage{hyphens}{url}
\usepackage[hidelinks]{hyperref}
\usepackage{graphicx}
\usepackage{amsmath}
\usepackage{authblk}
\usepackage{gensymb}
\usepackage{setspace}
\usepackage{amssymb}
\usepackage{tablefootnote}
\usepackage[table,xcdraw]{xcolor}
\usepackage{colortbl}
\usepackage{dirtytalk}

\usepackage[numbers,sort&compress]{natbib} % numeric citations, compressed ranges
\definecolor{low}{RGB}{198, 239, 206}
\definecolor{medium}{RGB}{255, 242, 204}
\definecolor{high}{RGB}{255, 199, 206}

\title{SQOUT: A Risk-Based Threat Analysis Framework for \\Quantum Communication Systems}
\author[1]{Michal Krelina\thanks{krelina@qudef.com}}
\author[2]{Tom Sorger}
\author[1]{Bob Dirks}
\affil[1]{QuDef B.V., Elektronicaweg 10, 2628XG Delft, The Netherlands}
\affil[2]{KTH Royal Institute of Technology, Brinellvägen 8, 114 28 Stockholm, Sweden}

\begin{document}
\maketitle

\begin{abstract}

Quantum communication systems, despite their theoretical security guarantees, face practical vulnerabilities across physical infrastructure, protocols, and classical subsystems that existing cybersecurity frameworks do not cover.
We propose a kill-chain-based threat model that organises quantum and classical Tactics, Techniques, and Procedures (TTPs) into end-to-end attack sequences, combined with an ISO/IEC 27005-compatible risk scoring methodology.
SQOUT, a threat-intelligence platform for quantum technologies, implements this approach and is used to analyse two concrete attack scenarios: Photon-Number Splitting (PNS) and detector-blinding attacks as case studies.
The risk assessment uses technique-level likelihood scoring with attack-tree product aggregation across kill-chain steps, producing governance-ready risk ratings for quantum-specific threat scenarios.

\end{abstract}

%\begin{keyword}
\textbf{Keywords:} Quantum communication, Quantum key distribution, QKD, Cybersecurity, Risk assessment, MITRE ATT\&CK, Kill chain, Photon-number splitting attack, SQOUT, Threat modelling
%\end{keyword}

%---------------------------------------------------------------------------
%---------------------------------------------------------------------------

\section{Introduction}\label{sec:intro}

Quantum Key Distribution (QKD) offers information-theoretically secure key exchange, with eavesdropping attempts detectable through quantum measurement disturbance. However, practical QKD implementations introduce vulnerabilities across physical infrastructure, protocols, and classical subsystems---attack surfaces that existing threat models do not address.

With quantum communication moving from experimental prototypes to operational deployments, driven by initiatives such as the European Quantum Communication Infrastructure (EuroQCI) \cite{euroqci_eu_commission}, understanding and managing their security risks becomes a practical necessity.
% Classical cybersecurity frameworks, such as MITRE ATT\&CK\footnote{\url{https://attack.mitre.org/}}, have proven effective in characterising adversarial behaviour in conventional IT environments. Inspired by this success, we aim to develop a similarly structured taxonomy and risk model tailored to the quantum domain.
Classical cybersecurity frameworks such as MITRE ATT\&CK \cite{mitre_attack}, the Lockheed Martin Cyber Kill Chain \cite{lockheed_cyber_kill_chain}, and NIST SP 800-30 \cite{nist80030r1} have proved effective for characterising adversarial behaviour in IT environments.
We adopt MITRE ATT\&CK as the structural basis because its technique-level granularity and continuous community updates make it directly actionable for detection and mitigation.
The taxonomy introduced here adapts the ATT\&CK conceptual structure to quantum communication systems rather than replicating it.

\subsection{Overview of Related Work}\label{subsec:related_work}

Prior works have characterised individual quantum attack vectors: detector-blinding attacks \cite{Lydersen_2010}, Trojan-horse exploits \cite{Vak2001}, CV-QKD side channels \cite{Huang_2013}, broader side-channel surveys \cite{BSI_2023}, and device-independent QKD security proofs \cite{Vazirani_Vidick_2014}. These studies analyse specific vulnerabilities but do not model the full end-to-end adversary path. QKD-focused risk assessments such as \cite{Sajeed_2021}, security analyses of QKD with realistic devices \cite{Xu_2020}, and surveys of quantum risk trends \cite{Brazaola-Vicario_2024} categorise threats and identify empirical risk patterns, yet they analyse individual attack surfaces rather than chaining reconnaissance, access, and exploitation into a single assessable scenario. Classical risk frameworks (NIST SP~800-30 \cite{nist80030r1}, ISO/IEC~27005 \cite{ISO_2022}) provide mature scoring methods but do not address quantum-specific attack mechanics. Our framework bridges this gap by building full kill chains that link quantum and classical techniques, scored using ISO/IEC~27005-compatible aggregation.

\subsection{Novelty of the current work}\label{subsec:contributions}

The novelty of the current work as well as its main contributions are as follows:
\begin{itemize}
  \item A kill-chain-based threat model that combines quantum and classical TTPs into end-to-end attack sequences, moving beyond isolated vulnerability catalogues.
  \item A quantitative risk scoring methodology using attack-tree product aggregation of technique-level likelihoods, aligned with ISO/IEC~27005 and compatible with standard risk matrices.
  \item SQOUT, a threat-intelligence platform that integrates the taxonomy, scoring, and interactive kill-chain analysis into a single operational tool.
\end{itemize}

\subsection{Paper Outline}

This paper introduces a structured framework for analysing threats to quantum communication systems by combining a taxonomy of attacks, kill-chain modelling, and quantitative risk evaluation. 
In section~\ref{sec:rta} we introduce the motivation for structured risk assessment in quantum communication deployments and discuss why classical cybersecurity frameworks must be adapted to account for quantum-specific vulnerabilities.
Section~\ref{sec:tax} follows with a taxonomy of attacks on quantum communication systems. The taxonomy classifies attacks according to objectives, mechanisms, deployment environments, and adversary characteristics, and establishes the conceptual vocabulary used throughout the remainder of the paper.
Then in section~\ref{sec:sqout}, we introduce SQOUT\footnote{Note an Open Access SQOUT with selected TTPs is available at \url{https://sqout.qudef.com/}.}, our threat-intelligence platform that operationalises the taxonomy. The platform provides a structured repository of quantum-specific Tactics, Techniques, and Procedures (TTPs) and enables the construction of adversarial kill chains that combine classical and quantum attack techniques.
Section~\ref{sec:risk_iso} builds on the kill-chain representation to perform quantitative risk assessment. Using an ISO/IEC 27005-aligned methodology, technique-level likelihood scores are aggregated into scenario-level risk ratings. Two case studies, Photon-Number Splitting (PNS) and detector-blinding attacks, illustrate the methodology.
Finally, in \autoref{sec:conclusions} we summarise the main findings and discuss future extensions of the framework, including conditional modelling of attack dependencies and integration with emerging quantum threat-intelligence sources.

%=========================================================
%=========================================================
%=========================================================
\section{Risk and Threat Assessment}\label{sec:rta}

QKD is often described as providing a theoretically unbreakable key-exchange mechanism for secure communication purposes. Efforts are ongoing to adopt it for high-security applications (e.g., critical infrastructure and classified governmental networks). Initiatives such as the EuroQCI aim to secure communications with QKD up to EU Secret level \cite{euroqci_conops_2024}. Similar programs in Asia, North America, and the private sector \cite{Gui2022,Li2025} underline the strategic importance of QKD for finance, energy, and defence, but also amplify the need for rigorous risk analyses, especially in the implementation and use of QKD systems.

Several National Security Agencies (NSAs) have warned of practical QKD vulnerabilities \cite{BSI_2023, position-2024}. The U.S. NSA highlights denial-of-service risks (via induced eavesdropping alarms), hardware flaws exploited in lab attacks, and insider threats from trusted relays \cite{NSA}. European bodies (BSI, ANSSI, AIVD, etc.) similarly note the immature security maturity of QKD, high infrastructure costs and niche applicability, recommending parallel investment in post-quantum cryptography (PQC) \cite{position-2024} for a more scalable defence.

To systematically analyse these risks, a structured description of possible attacks is required. The following section introduces a taxonomy of attacks on quantum communication systems that provides the foundation for the threat model used in SQOUT.

%=========================================================
%=========================================================
%=========================================================
\section{Taxonomy of Attacks on Quantum Communication Systems }\label{sec:tax}

A risk and threat assessment for quantum communications must cover:

\begin{itemize}
  \item \textbf{Physical infrastructure:} Fibre, satellite, and device risks (tampering, loss, environment).
  \item \textbf{Protocol threats:} Quantum-specific attacks (photon-number splitting, Trojan horse, side channels).
  \item \textbf{Classical interdependence:} Weak links in hybrid quantum-classical deployments.
  \item \textbf{Operational factors:} Human error, supply chain flaws, and insider risk.
\end{itemize}

A key novelty of our work is that, beyond cataloguing individual quantum and classical techniques, we organise them into sequential kill-chain phases. A kill chain (sometimes also called a risk scenario) represents an ordered sequence of adversarial techniques that must succeed for an attack scenario to be realised. This end-to-end view captures inter-step dependencies and enables holistic risk scoring.

This section defines and categorises attacks on quantum communication systems, focusing on their objectives, mechanisms, and contexts. By aligning with classical Tactics, Techniques, and Procedures (TTPs) frameworks,
%\footnote{\url{https://www.exabeam.com/explainers/what-are-ttps/what-are-ttps-and-how-understanding-them-can-help-prevent-the-next-incident/}}
 we aim to make quantum-specific threats accessible and actionable for security professionals.

%==============================================
%==============================================
\subsection{Defining Attacks}

\begin{figure}
    \centering
    \includegraphics[width=0.7\linewidth]{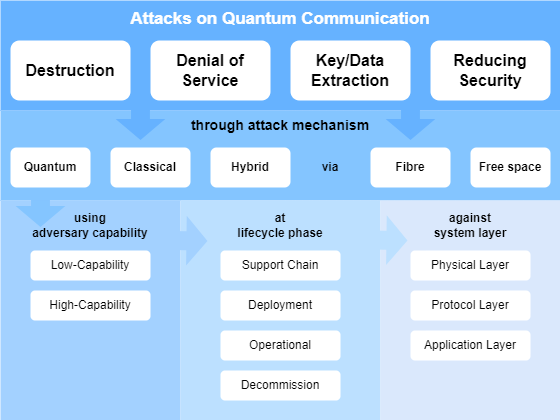}
    \caption{Hierarchical taxonomy of attacks on quantum communication systems.}
    \label{fig:taxonomy}
\end{figure}

We categorise attacks by their objectives (\autoref{fig:taxonomy}), reflecting the adversary's intent and the potential impact on the system:

\begin{itemize}
    \item \textbf{Destruction:}
    Permanent and non-reversible compromise of quantum communication systems until repaired or replaced.
    Destruction can be categorised into two types:
    \begin{itemize}
        \item \textbf{Physical Destruction:}
        Direct physical damage to infrastructure or components, such as cutting optical fibres, burning out single-photon detectors, tampering with cryogenic equipment, or damaging optical hardware.
        \item \textbf{Logical/Software Destruction:}
        Permanent corruption or deletion of critical control, calibration, or key management software, rendering the system inoperable even though the hardware remains intact.
    \end{itemize}
    Both forms result in long-term loss of service and require repair or replacement to restore operation.
    
    \item \textbf{Denial of Service (DoS):}
    Temporary and reversible disruption of quantum communication functionality by preventing legitimate use of system resources or communication channels.
    Examples include jamming free-space optical links, saturating detectors, or overloading quantum channels to impede key generation or transmission.
    Unlike destruction, DoS attacks cease to have effect once the interference or resource exhaustion stops.

    \item \textbf{Quantum Key or Data Extraction:} Involving attempts to intercept encryption keys generated via QKD or extract sensitive quantum data (of quantum communication services beyond QKD, e.g., blind quantum computing). This category includes attacks aiming for \textit{full key/data extraction}, where the entire quantum key or dataset is compromised, and \textit{partial key/data extraction}, where only fragments are obtained but could still pose significant security risks. Examples of these quantum-specific attacks include photon-number-splitting (PNS) \cite{Brassard_Lütkenhaus_Mor_Sanders_2000} or Trojan-horse attacks \cite{Vakhitov_Makarov_Hjelme_2001}.

    \item \textbf{Reducing Security:} Attacks involve introducing weaknesses to compromise a system or reduce the security parameter below the threshold guaranteed by its quantum security proof. Examples include laser damage attacks on the watchdog or attenuator \cite{Makarov_2016}, or laser seeding attacks \cite{Huang_2019}.
\end{itemize}

%==============================================
%==============================================
\subsection{Attack Mechanism}

In addition to defining attack objectives, we classify attacks according to their dominant mechanisms and operational domains, acknowledging that most practical threats to QKD systems span both quantum and classical layers:

\begin{itemize}
    \item \textbf{Quantum-Dominant Attacks:}
    Attacks primarily exploiting imperfections in quantum state preparation, transmission, or measurement processes.
    These rely on manipulating quantum phenomena—such as photon statistics, detector timing, or entanglement correlations—although classical information exchange (e.g., sifting data) may still be required to complete the exploit.
    Examples include photon-number-splitting and time-shift attacks.

    \item \textbf{Classical-Dominant Attacks:}
    Attacks that mainly target the classical subsystems supporting QKD operation, such as control software, synchronization, authentication, or key management.
    Examples include injecting false timing signals, tampering with error-correction routines, or compromising authentication mechanisms.

    \item \textbf{Cross-Layer (Hybrid) Attacks:}
    Coordinated, multi-domain exploits that deliberately combine quantum and classical manipulations to achieve outcomes unattainable by either layer alone.
    For instance, compromising calibration software to induce detector mismatches subsequently exploited by a faked-state quantum attack~\cite{Makarov_2009}, or altering authentication routines to enable a man-in-the-middle quantum intercept-resend.
\end{itemize}

We also classify attacks by their deployment environment:

\begin{itemize}
    \item  \textbf{Fibre-Based Attacks:} These target vulnerabilities in the optical fibre infrastructure, such as tapping, bending-induced signal leakage, or physical intercept or sabotage. For example, a naive intercept-and-resend attack is relatively straightforward to perform on fibre, as the medium offers stable and controlled transmission conditions.

    \item \textbf{Free-Space Attacks:} These focus on free-space communication channels, such as satellite links or ground-to-satellite connections. While similar attack concepts may apply, free-space environments present unique challenges, such as atmospheric interference and line-of-sight constraints, making even basic attacks, like intercept-and-resend, more complex.
    
\end{itemize}

The differences in attack feasibility between fibre and free-space deployments require environment-specific security measures.

%==============================================
%==============================================
\subsection{Additional Classifications}

In addition to objectives, mechanisms, and environments, we further distinguish attacks along several key dimensions:  

\textbf{Adversary Capabilities:} from low-capability opportunists with basic tools (e.g., simple fibre taps) to nation-state actors with advanced quantum hardware and deep R\&D backing. 

\textbf{Attack Phase:} spanning the supply-chain (hardware trojans, firmware tampering), deployment (intercepting shipments, misconfiguration), operational (intercept-and-resend, side-channel exploits), and decommissioning (recovering residual quantum data) stages.  

\textbf{Target System Layers:} from the physical layer (optical-fibre tapping, free-space jamming) through the protocol layer (man-in-the-middle, information-reconciliation side channels) to the application layer (compromising quantum-secured services or post-quantum transitions).

%==============================================
%==============================================
\subsection{Tactics}

Building on MITRE ATT\&CK and D3FEND \cite{mitre_d3fend}, we define offensive and defensive tactics tailored to quantum communications. The meaning of each tactic remains unchanged; it is simply applied in the quantum context to guide security practitioners.

Certain techniques, e.g., PNS, span multiple tactics (execution, collection, exfiltration). Rather than force one category, we assign them to all relevant tactics, maintaining a flexible, comprehensive threat model.

%=========================================================
%=========================================================
%=========================================================
\section{SQOUT}\label{sec:sqout}

Building on the attack taxonomy defined in \autoref{sec:tax}, we introduce SQOUT, QuDef's \cite{qudef} threat-intelligence platform that organises quantum attack techniques into a structured, queryable knowledge base. It covers quantum communication, computing, and sensing, and follows the MITRE ATT\&CK organisational model. The platform maintains a hierarchical repository of adversarial techniques and defensive countermeasures, mapped to quantum protocols, detection methods, hardware elements, software modules, and threat actors.

SQOUT provides a matrix of TTPs covering quantum-specific compromise scenarios. For each technique, it records descriptions, indicators of compromise, and countermeasures.

In its current form, SQOUT is a prototype demonstrating the feasibility of structured quantum threat intelligence. It has not undergone formal certification and remains under active development, while already supporting a number of security analyses.

Beyond its static knowledge base, SQOUT includes a suite of interactive applications that support the end-to-end threat-modelling lifecycle. Users can graphically build a quantum system architecture, annotate it with relevant TTPs, and execute automated risk assessments or \say{what if} analyses. These tools support identification of high-impact vulnerabilities and validation of proposed defences.

%==============================================
%==============================================
\subsection{Kill Chain Example}\label{sec:pns_killchain_example}
\begin{figure}[htb!]
    \centering
    \includegraphics[width=0.9\linewidth]{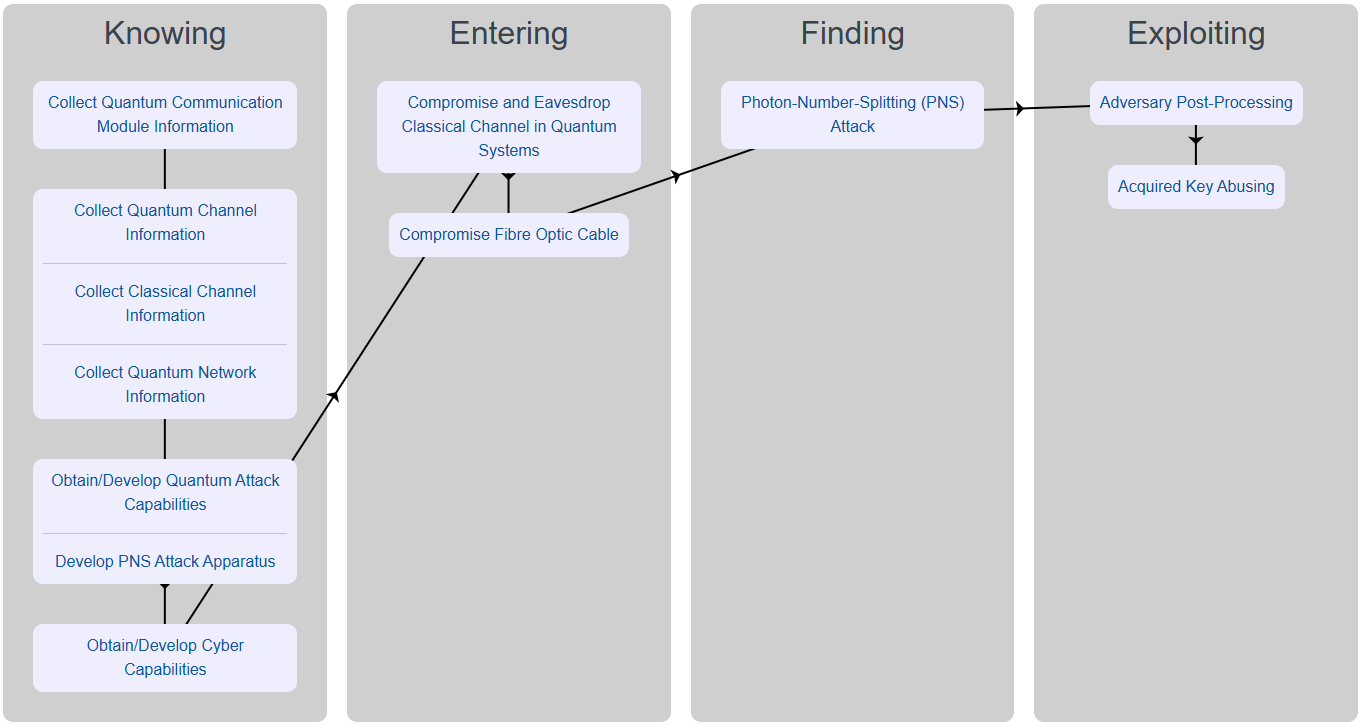}
    \caption{PNS attack kill chain. Extracted from SQOUT.}
    \label{fig:pns-killchain}
\end{figure}

As an example, the PNS \cite{Brassard_Lütkenhaus_Mor_Sanders_2000} attack kill chain illustrates the step-by-step progression of a threat actor targeting a quantum communication system, divided into four phases: Knowing, Entering, Finding, and Exploiting, as visualised in \autoref{fig:pns-killchain}.

\textbf{Knowing}: The initial phase focuses on gathering critical information about the quantum communication system. Threat actors collect details about the quantum communication module (protocols, sources, properties) and the supporting quantum and classical channels, including their locations. Concurrently, they develop specialised quantum attack capabilities, such as a PNS attack apparatus requiring non-demolition measurement and quantum memory, along with the cyber tools necessary to extract information from classical links.

\textbf{Entering}: In this phase, the adversary begins compromising system components to gain entry. This includes eavesdropping on the classical channel used in the quantum communication system and physically accessing the fibre optic cable carrying the quantum link. These steps enable the attacker to position themselves for active engagement with the quantum communication.

\textbf{Finding}: The attacker executes the PNS attack itself, exploiting vulnerabilities in multi-photon pulses to extract quantum key information. This step represents the core malicious activity and requires precise quantum attack tools and expertise.

\textbf{Exploiting}: In the final phase, the adversary processes the intercepted quantum data to extract a key identical to that of the legitimate communication parties. They then utilise or abuse the acquired key, potentially leading to unauthorised access or further exploitation of the system.

This example illustrates how TTPs from the SQOUT catalogue compose into a complete kill chain that can be scored using the risk methodology in the next section.

Quantitative risk assessment of these kill chains requires a scoring methodology, which the following section introduces.

%=========================================================
%=========================================================
%=========================================================
\section{Methodology: Risk Analysis and Evaluation}\label{sec:risk_iso}

This section presents the methodological core of the paper: a reproducible procedure for converting a kill chain into an ISO/IEC~27005-compliant risk rating. The methodology is fully specified by (i)~the technique-level scoring rules of \autoref{subsec:risk_analysis}, (ii)~the aggregation operators and discretisation bounds defined in the same subsection, and (iii)~the global modifier and matrix lookup of \autoref{subsec:risk_evaluation}. The two worked case studies in \autoref{sec:pns_example} and \autoref{sec:blinding_example} act as demonstrated proofs of use, executing every methodological step end to end on two independent attack scenarios with materially different resource profiles.

We now apply this ISO/IEC 27005-compliant risk assessment \cite{ISO_2022} to the kill-chain structures and SQOUT technique inventory introduced above. ISO/IEC 27005 structures risk management into (a) context establishment, (b) risk analysis (likelihood and impact estimation), and (c) risk evaluation (matrix lookup), followed by treatment and review.
Our model retains likelihood estimates at the technique level and aggregates them into a single risk rating per scenario.

%==============================================
%==============================================
\subsection{Context Establishment}  
Prior to numeric scoring, the organisation must define the following properties:
\begin{itemize}
  \item \emph{Scope:} the boundaries of the quantum‐communication system under review (e.g., the free-space QKD link, its repeater nodes, and classical control channels).
  \item \emph{Risk acceptance criteria:} the thresholds in the 5×5 matrix that delineate Acceptable, Tolerable, and Unacceptable risk.
  \item \emph{Roles and responsibilities:} assignment of who conducts analysis, who reviews and signs off residual risk, and who implements countermeasures.
\end{itemize}
These elements ensure that subsequent likelihood and impact estimates are judged against pre‐agreed organisational objectives and compliance requirements.

%==============================================
%==============================================
\subsection{Risk Analysis}\label{subsec:risk_analysis}

In this model, a kill chain refers to a sequence of adversarial steps (techniques) required to execute a complete attack, from reconnaissance through exploitation. A risk scenario is defined as the successful completion of a specific kill chain under a given set of system conditions and controls. Risk is evaluated at the scenario level, with likelihood derived from the individual steps and impact representing the consequence of full scenario success.

The risk analysis produces two separate ordinal values on a scale of 1 to 5 for each \emph{risk scenario} (a complete kill chain, such as a PNS attack on a QKD link using a weak coherent pulse source without decoy-state protection):  
\[
\text{Likelihood }L\in\{1,\dots,5\},
\quad
\text{Impact }I\in\{1,\dots,5\}.
\]

%==============================================
\subsubsection{Likelihood Estimation}\label{subsub:likelihood}

Likelihood reflects the probability that an adversary successfully executes a complete attack scenario (i.e., the entire kill chain) given the existing system controls. In the proposed model, likelihood is first estimated at the technique level. Each technique \(i\) in the kill chain is assigned two base scores: a Threat score \(T_i\) that reflects the capability class of adversary that would attempt and succeed at the technique, and an Exposure score \(E_i\) that represents the target system component's combined accessibility and vulnerability to that technique.

These scores are combined to produce an intermediate likelihood contribution for each technique. The individual contributions are subsequently aggregated across the kill chain to estimate the scenario-level likelihood of a successful attack.

The impact \(I\) of the attack is assessed separately at the scenario level and reflects the consequences if the entire kill chain succeeds.

\paragraph{Technique-level Scoring.}
For each step \(i\) in the kill chain (as enumerated in \autoref{sec:pns_killchain_example}), assign (reference scales and modifier examples are given in \autoref{tab:tv_scores} and \autoref{tab:modifiers})  
\begin{equation}
T_i,\;E_i\;\in\{1,2,..,n_{max}\},
\qquad
m_i\in(0,\mu_{\max}\rangle
\end{equation} 
where \(T_i\) is the threat score (adversary capability class), \(E_i\) the exposure score (combined accessibility and vulnerability), and we use \(n_{max}=5\). The parameter \(m_i\) is a technique‐specific multiplier (e.g., extra countermeasure or environmental hindrance) which is bounded by  \(\mu_{\max}=2\)  to ensure that contextual modifiers adjust the base likelihood without dominating it. This helps to preserve the interpretability of the base Threat and Exposure scores while still allowing moderate amplification or reduction due to environmental factors or countermeasures.  Define the step’s raw likelihood contribution:  
\begin{equation}\label{eq:step_likelihood}
    \ell_i = (T_i\times E_i)\times m_i.
\end{equation}
The resulting value \(\ell_i\) is not yet a probability but a relative likelihood score used for subsequent aggregation. 

\paragraph{Aggregation to Scenario Raw Likelihood.}  
The individual step contributions \(\{\ell_i\}\) must be combined into a single continuous measure \(L_{\rm raw}\) that reflects the ease with which an adversary can complete the entire kill chain.  Let \(N\) be the total number of steps (techniques) in that kill chain.  Three common aggregation strategies are:

\begin{itemize}
  \item \textbf{Maximum:}  
    \(L_{\rm raw}=\max_i \ell_i\). This \say{worst‐step} approach assumes that the easiest step for the attacker dominates the scenario likelihood.  It highlights the single weakest link in the chain and is conservative when any one step could enable full compromise.
  
  \item \textbf{Average:}
    \(L_{\rm raw}=\frac{1}{N}\sum_i \ell_i\). By computing the mean of all step contributions, this method treats each phase as equally important.  It smooths out extreme values and is appropriate when partial difficulty in one step can be offset by ease in another.  Note, however, that the average can overestimate feasibility for chains containing one near-infeasible step, because it does not reflect the AND-logic of sequential attack success.
  
\item \textbf{Probabilistic (attack-tree product):}
  First, convert each step’s likelihood into a success probability:
\begin{equation}
    p_i \;=\;\min\!\Bigl(1,\;\frac{\ell_i}{T_{\max}\cdot E_{\max}}\Bigr) = \min\!\Bigl(1,\;\frac{\ell_i}{25}\Bigr),
\end{equation}

where \(T_{\max}\) and \(E_{\max}\) denote the maximum values of the Threat and Exposure scales. The denominator $T_{\max}\cdot E_{\max}$ represents the maximum possible base score (before the technique multiplier $m_i$). When $m_i > 1$, the raw product $\ell_i = T_i \cdot E_i \cdot m_i$ can exceed $T_{\max}\cdot E_{\max}$.
Values exceeding this bound are intentionally clamped to 
via $\min(1,\cdot)$, reflecting that favourable conditions make the step almost certain once the attack is attempted.
 
Assuming conditional independence of step success probabilities (see the ``Independence Assumptions'' paragraph below for discussion and limitations), the chain success probability is the product over all steps (AND-node in attack-tree terminology):
\begin{equation}
    P_{\rm chain} \;=\;\prod_{i=1}^N p_i.
\end{equation}
  $P_{\rm chain}$ is the compound probability that the attacker succeeds at every step --- the standard aggregation for a serial attack chain in attack-tree analysis \cite{Schneier1999}, Bayesian attack graphs \cite{NISTIR7788,Wang2007}, and network vulnerability analysis \cite{Xie2010}.

    Unlike the maximum and average methods, the attack-tree product does not produce a score-valued $L_{\rm raw}$; its output is a probability.  The global multiplier $M$ is therefore applied in log-space via exponentiation, which is the natural analogue of linear scaling on a logarithmic axis:
    \begin{equation}
        P_{\rm adj} \;=\; P_{\rm chain}^{\,1/M}.
    \end{equation}
    Since $\log_{10} P_{\rm adj} = \log_{10} P_{\rm chain}\,/\,M$, a higher~$M$ (more threatening environment) compresses the negative log-probability, increasing the adjusted probability, while a lower~$M$ (hardened environment) stretches it, reducing the adjusted probability. For $M=1$ the adjustment is neutral.  $P_{\rm adj}$ is then discretised via the logarithmic bins in \autoref{tab:log_bins}.

\end{itemize}

\autoref{tab:agg_guidance} summarises how to select the aggregation method based on system criticality, risk appetite, and compliance considerations.

\begin{table}[htb!]
\centering
\small
\begin{tabular}{|p{6cm}|p{3cm}|p{6cm}|}
\hline
\textbf{Context / Requirement} & \textbf{Aggregation Method} & \textbf{Rationale} \\\hline
Safety‐ or life‐critical systems & Max & Ensures no single weak step is overlooked (upper‐bound risk) \\\hline
Statutory/regulatory compliance & Max or Attack‐tree product & Max for conservative compliance; product for realistic probabilistic traceability \\\hline
Balanced risk appetite & Attack‐tree product & Reflects sequential success probabilities in a series‐system model \\\hline
Early‐stage or resource‐limited assessments & Average & Smoothed estimate of per-step difficulty for resource planning; limited discrimination between scenarios due to averaging compression --- not recommended as a primary risk indicator \\\hline
\end{tabular}
\caption{Guidance for choosing an aggregation method based on system criticality, risk appetite, and compliance needs.}
\label{tab:agg_guidance}
\end{table}

\paragraph{Theoretical Rationale for the Attack-Tree Product.}
In the attack-tree analysis \cite{Schneier1999}, a kill chain corresponds to an AND-decomposition: the attacker must succeed at every step for the full chain to succeed.  Under the standard independence assumption, the compound success probability is the product of step probabilities.  This is the same aggregation used in Bayesian attack graphs, where AND-nodes compute $P(\text{goal})=\prod p_i$ \cite{NISTIR7788,Wang2007,Xie2010}, and in series-system reliability models, where overall system success probability equals the product of component success probabilities \cite{Modarres2016}.  
Because multi-step chain probabilities span many orders of magnitude, the attack-tree product requires logarithmic discretisation (defined in the paragraph \emph{Logarithmic Aggregation of Attack-Tree Probabilities} below) to enable meaningful discrimination between chains of different feasibility.

\paragraph{Ordinal Scales and Arithmetic Operations.}
The Threat and Exposure scores are ordinal, meaning that the intervals between adjacent values (e.g., $T=2$ vs.\ $T=3$) are not guaranteed to be equal in a strict measurement-theoretic sense. Performing arithmetic operations such as multiplication and product-based aggregation on ordinal data is, strictly speaking, an approximation. However, this practice is standard in widely adopted risk frameworks, including ISO/IEC~27005 \cite{ISO_2022} and FAIR \cite{Jones2006FAIR}, where ordinal likelihood and impact scales are routinely multiplied to produce risk ratings. The final discretisation step (defined in the paragraph \emph{Global Adjustment and Discretisation} below) maps the continuous intermediate result back to an ordinal category ($L \in \{1,\dots,5\}$), so the output remains on the same ordinal scale as the inputs. The intermediate arithmetic is therefore a modelling convenience that enables structured aggregation across kill-chain steps, not a claim of interval-scale precision. Subjectivity in the assignment of $T$ and $E$ values is inherent to any expert-driven risk framework; consistency is ensured by using the reference definitions in \autoref{tab:tv_scores} and applying them uniformly across all scenarios.

\paragraph{Independence Assumptions.}
The attack-tree product aggregation assumes that kill-chain steps succeed independently. In practice, steps are often positively correlated, for example, physical access to a fibre link may enable both classical eavesdropping and quantum-channel interception. Under positive correlation, the true compound probability of full-chain success is \emph{higher} than the product of marginal probabilities. Consequently, the independence assumption yields a \emph{lower bound} on scenario likelihood, making the attack-tree product estimate conservative rather than optimistic. Extending the model to conditional probabilities (e.g., via a Bayesian network where each node's success probability depends on its predecessors) is straightforward within the framework and is planned as future work. It is worth noting that similar independence assumptions are standard at the base level of established risk frameworks, including NIST SP~800-30 \cite{nist80030r1} and classical kill-chain models, where step-level likelihoods are typically assessed and combined without explicit conditional modelling.

\paragraph{Global Adjustment and Discretisation.}  
Apply an environment-level multiplier \(M \in (0,2\rangle\) which globally adjusts likelihoods to reflect the overall threat context in which the kill chains are assessed. For example, a higher $M$ (e.g., 1.5) may represent a highly capable, state-sponsored adversary or a critical national infrastructure deployment, while a lower $M$ (e.g., 0.7) could reflect strong system hardening or low adversary motivation.

For the maximum and average aggregation methods, $M$ scales the raw score linearly:
\begin{equation}
L_{\rm adj}=L_{\rm raw}\times M.
\end{equation}
For the attack-tree product, $M$ acts in log-space via $P_{\rm adj} = P_{\rm chain}^{1/M}$ (see above), ensuring that the global multiplier has proportional impact on the logarithmic discretisation scale.  Because the product method operates on a logarithmic scale while the max and average methods operate on a linear scale, the same $M$ value produces different absolute magnitudes of adjustment: for example, $M=1.5$ shifts $P_{\rm adj}$ by roughly one logarithmic bin width (${\sim}2$ orders of magnitude), which corresponds to approximately one ordinal level---comparable to the ${\sim}1$-level shift produced by linear scaling of the max or average raw scores.

The discretisation bounds are fixed to the base score range, independent of the scenario set:
\begin{equation}
L_{\min}=0,\qquad L_{\max}=T_{\max}\cdot E_{\max}=n_{\max}^2.
\end{equation}
For our 1--5 scales, $L_{\max}=25$.  Divide $[L_{\min},\,L_{\max}]$ into five equal intervals and assign the ordinal likelihood
\begin{equation}
L=1+\left\lfloor 5\,\frac{L_{\rm adj}-L_{\min}}{L_{\max}-L_{\min}}\right\rfloor,
\end{equation}
clamped to $\{1,\dots,5\}$.

The discretisation bounds $L_{\min}=0$, $L_{\max}=T_{\max}\cdot E_{\max}$ are consistent with the attack-tree product normalisation, which uses the same denominator $T_{\max}\cdot E_{\max}$ to convert step scores to probabilities. Each bin therefore corresponds to a 20\% step-probability range: $L=1$ maps to $p<0.2$ (hard step), $L=5$ to $p\ge0.8$ (near-certain step).  Because the bounds derive from the scale definition alone, they are deterministic: adding or removing scenarios from the evaluation set never changes existing discretised values.

\paragraph{Effect of Modifiers on the Discretisation Bounds.}
Technique modifiers $m_i \neq 1$ and the global multiplier $M \neq 1$ shift intermediate values outside the base score range $[0,\;T_{\max}\cdot E_{\max}]$.  The framework handles both directions consistently across all three aggregation methods.

\emph{Amplification ($m_i > 1$ or $M > 1$).}
When modifiers push values above the base maximum $T_{\max}\cdot E_{\max}$:
\begin{itemize}[nosep]
  \item \emph{Attack-tree product:} $p_i = \min(1,\;\ell_i / (T_{\max}\cdot E_{\max}))$ clamps to $p_i = 1$, treating the step as certain to succeed.  The remaining steps still contribute their probabilities to the chain product.
  \item \emph{Maximum and average:} The discretisation formula may yield a value above~$5$, which is clamped to $L=5$.
\end{itemize}
The amplifying modifier still has an effect when the unadjusted score is below the base maximum --- for example, $\ell_i = 3\times4\times1.2 = 14.4$ (below~25, no clamping) versus $\ell_i = 5\times5\times1.2 = 30$ (above~25, clamped to $p_i=1$ or $L=5$).  If many steps in a kill chain are amplified to the ceiling, the base $T$ and $E$ scores should be re-examined rather than relying on large modifiers.

\emph{Mitigation ($m_i < 1$ or $M < 1$).}
When modifiers reduce values toward the bottom of the scale:
\begin{itemize}[nosep]
  \item \emph{Attack-tree product:} A mitigated step receives a lower $p_i$, which directly reduces the chain product $P_{\rm chain}$.  For example, applying $m_i=0.2$ to a step with $T_i=4$, $E_i=4$ yields $\ell_i = 3.2$ and $p_i = 0.128$, compared to $p_i = 0.64$ without the modifier --- a $5\times$ reduction in the step's success probability.  A global modifier $M < 1$ acts via $P_{\rm adj} = P_{\rm chain}^{1/M}$; since $1/M > 1$ and $P_{\rm chain} < 1$, the adjusted probability $P_{\rm adj} < P_{\rm chain}$, correctly reducing the scenario likelihood.
  \item \emph{Maximum and average:} $L_{\rm adj} = L_{\rm raw}\times M$ decreases proportionally.  The discretisation formula naturally maps small values to $L=1$.  No lower clamping beyond $L=1$ is needed, since $L_{\min}=0$ and $L_{\rm adj}$ remains non-negative.
\end{itemize}
Note that mitigation modifiers are not symmetric to amplification in their impact on the ordinal scale: $m_i = 0.2$ can reduce a step's contribution by $5\times$, while $m_i = \mu_{\max} = 2$ can at most double it (before clamping).  This asymmetry is intentional (see \autoref{subsec:tv_and_mods}): a targeted countermeasure can near-eliminate a specific step, whereas aggravating factors are bounded more tightly because the baseline $T$ and $E$ scores already capture the primary attack feasibility.

\paragraph{Logarithmic Aggregation of Attack-Tree Probabilities.}
For the attack-tree product, the adjusted probability $P_{\rm adj}$ spans many orders of magnitude, making linear binning unsuitable.  We discretise on a logarithmic scale, assigning ordinal likelihood levels based on order-of-magnitude bands.

The bin boundaries are defined as absolute success-probability thresholds, analogous to the frequency bands used in probabilistic risk assessment (e.g., core-damage frequency bands in NUREG-1150 \cite{NUREG1150} and likelihood categories in the NASA PRA Procedures Guide \cite{NASA_PRA_2011}).  To calibrate the boundaries, consider the theoretical range of $P_{\rm adj}$ for an $N$-step chain with step probabilities $p_i \in [p_{\min},\;1]$, where $p_{\min} = \ell_{\min}/(T_{\max} \cdot E_{\max})$.  The log-scale range is $[\,N\log_{10} p_{\min}/M,\;0\,]$.  For the kill chains in this paper ($N=9$, $M=1$), taking the smallest modifier value $m_i=0.2$ with the minimum base scores $T_i=E_i=1$ gives $\ell_{\min}=1\times1\times0.2=0.2$ and thus $p_{\min}=0.2/25=0.008$, yielding a log-scale range of approximately $[-19,\;0]$.  We select five bins of equal width $\Delta = 2$ orders of magnitude, anchored so that the upper boundary ($\log_{10} P = 0$, i.e.\ $P=1$) corresponds to $L=5$:

\begin{table}[htb!]
\centering
\begin{tabular}{|c|l|l|l|}
\hline
$L$ & \textbf{Label} & $P_{\rm adj}$ \textbf{range} & $\log_{10}(P_{\rm adj})$ \textbf{range} \\\hline
1 & Very unlikely & $P_{\rm adj} < 10^{-8}$ & $(-\infty,\;-8]$ \\\hline
2 & Unlikely & $10^{-8}\le P_{\rm adj} < 10^{-6}$ & $(-8,\;-6]$ \\\hline
3 & Possible & $10^{-6}\le P_{\rm adj} < 10^{-4}$ & $(-6,\;-4]$ \\\hline
4 & Likely & $10^{-4}\le P_{\rm adj} < 10^{-2}$ & $(-4,\;-2]$ \\\hline
5 & Frequent & $P_{\rm adj} \ge 10^{-2}$ & $(-2,\;0]$ \\\hline
\end{tabular}
\caption{Logarithmic discretisation of the globally adjusted attack-tree probability $P_{\rm adj} = P_{\rm chain}^{1/M}$ into ordinal likelihood levels.  The five bins of width $\Delta=2$ are anchored at $\log_{10} P_{\rm adj} = 0$; values falling below $10^{-8}$ are collected into the $L=1$ overflow bin.}
\label{tab:log_bins}
\end{table}

The four resolved bins ($L=2$--$5$) cover eight orders of magnitude; scenarios with $P_{\rm adj} < 10^{-8}$ are collected into the $L=1$ overflow bin.  Because the bins represent absolute success probabilities, they are chain-length-independent: a longer chain with more steps naturally produces a lower $P_{\rm adj}$, correctly reflecting the greater compound difficulty.  For assessments involving substantially shorter or longer chains (e.g., $N \le 4$ or $N \ge 15$), the bin width $\Delta$ or the number of resolved bins may be adjusted to match the effective dynamic range, following the same anchoring principle.

The maximum and average aggregation methods produce values on a linear scale and continue to use the linear discretisation above.  The attack-tree product produces values on a multiplicative (logarithmic) scale and therefore requires logarithmic binning.  Each discretisation is matched to the mathematical nature of its aggregation method.

%==============================================
\subsubsection{Impact Estimation }\label{subsub:impact}

Impact \(I\) represents the severity of the consequence if the entire kill-chain scenario succeeds. Following ISO/IEC~27005, impact is assessed as a single ordinal value $I \in \{1,\dots,5\}$, ranging from very low ($I=1$, negligible operational effect) through medium ($I=3$, significant service degradation) to very high ($I=5$, catastrophic compromise of mission-critical assets). In practice, one chooses \(I\) based on the worst-case effect of the scenario (e.g., potential data loss, service outage, or national security implications) without indexing by step. For example, if a PNS attack on a QKD link would expose mission-critical keys, one might assign \(I=5\).

%==============================================
%==============================================
\subsection{Risk Evaluation}\label{subsec:risk_evaluation}

With Likelihood \(L\) and Impact \(I\) on matching 1-5 scales, the final risk rating \(R = L \times I\in\{1,\dots,25\}\) is obtained from the standard ISO/IEC 27005 matrix:
\begin{equation}
R = \mathrm{Matrix}[L,I].
\end{equation}

\begin{table}[htb!]
\centering
\rowcolors{2}{white}{gray!10}
\begin{tabular}{c|
  >{\centering\arraybackslash}p{1.8cm}
  >{\centering\arraybackslash}p{1.8cm}
  >{\centering\arraybackslash}p{1.8cm}
  >{\centering\arraybackslash}p{1.8cm}
  >{\centering\arraybackslash}p{1.8cm}
}
   & \shortstack{\textbf{$I=1$}\\\textbf{Very low}} 
   & \shortstack{\textbf{$I=2$}\\\textbf{Low}} 
   & \shortstack{\textbf{$I=3$}\\\textbf{Medium}} 
   & \shortstack{\textbf{$I=4$}\\\textbf{High}} 
   & \shortstack{\textbf{$I=5$}\\\textbf{Very high}}\\\hline
\shortstack{\textbf{$L=1$}\\\textbf{Very unlikely}} 
  & \cellcolor{low}\shortstack{$1$\\ (Low)} 
  & \cellcolor{low}\shortstack{$2$\\ (Low)}
  & \cellcolor{low}\shortstack{$3$\\ (Low)} 
  & \cellcolor{medium}\shortstack{$4$\\ (Medium)} 
  & \cellcolor{medium}\shortstack{$5$\\ (Medium)} \\
\shortstack{\textbf{$L=2$}\\\textbf{Unlikely}} 
  & \cellcolor{low}\shortstack{$2$\\ (Low)} 
  & \cellcolor{low}\shortstack{$4$\\ (Low)} 
  & \cellcolor{medium}\shortstack{$6$\\ (Medium)}  
  & \cellcolor{medium}\shortstack{$8$\\ (Medium)}    
  & \cellcolor{medium}\shortstack{$10$\\ (Medium)} \\
\shortstack{\textbf{$L=3$}\\\textbf{Possible}} 
  & \cellcolor{low}\shortstack{$3$\\ (Low)}   
  & \cellcolor{medium}\shortstack{$6$\\ (Medium)}   
  & \cellcolor{medium}\shortstack{$9$\\ (Medium)}   
  & \cellcolor{medium}\shortstack{$12$\\ (Medium)}   
  & \cellcolor{high}\shortstack{$15$\\ (High)} \\
\shortstack{\textbf{$L=4$}\\\textbf{Likely}} 
  & \cellcolor{medium}\shortstack{$4$\\ (Medium)}     
  & \cellcolor{medium}\shortstack{$8$\\ (Medium)}      
  & \cellcolor{medium}\shortstack{$12$\\ (Medium)}     
  & \cellcolor{high}\shortstack{$16$\\ (High)} 
  & \cellcolor{high}\shortstack{$20$\\ (High)}\\
\shortstack{\textbf{$L=5$}\\\textbf{Frequent}} 
  & \cellcolor{medium}\shortstack{$5$\\ (Medium)}      
  & \cellcolor{medium}\shortstack{$10$\\ (Medium)}      
  & \cellcolor{high}\shortstack{$15$\\ (High)}     
  & \cellcolor{high}\shortstack{$20$\\ (High)} 
  & \cellcolor{high}\shortstack{$25$\\ (High)}\\
\end{tabular}
\caption{Risk rating matrix with numeric and descriptive values.}
\label{tab:risk_matrix}
\end{table}

The risk categories in \autoref{tab:risk_matrix} are defined by the following boundaries: Low ($R \in \{1,2,3\}$), Medium ($R \in \{4,5,6,8,9,10,12\}$), and High ($R \in \{15,16,20,25\}$).
Organisations then compare \(R\) to their risk‐acceptance criteria (e.g., treat all \(R\ge 15\) as High) and plan mitigation accordingly.

%==============================================
%==============================================
\subsection{Base Likelihood Scoring and Multiplicative Modifiers}\label{subsec:tv_and_mods}

Each attack technique \(i\) is assigned base scores $T_i$ and $E_i$ denoting, respectively, the adversary capability class required to attempt and succeed at the technique, and the target component's combined accessibility and vulnerability, both assessed for a \say{typical} environment (moderate controls, average adversary). Higher $T$ increases likelihood because a more capable adversary---the kind that would attempt a high-$T$ technique---is also more likely to succeed. \autoref{tab:tv_scores} provides reference definitions.

\begin{table}[htb!]
\centering
\begin{tabular}{|c|p{12cm}|}
\hline
\textbf{Threat (\(T\))} & \textbf{Definition} \\ \hline
1 & Opportunistic adversary; minimal capability needed and widely available (e.g., severing an exposed fibre). \\ \hline
2 & Moderately skilled adversary with basic tools and hacking or optical expertise. \\ \hline
3 & Advanced adversary with classical cyber-attack capability or partial quantum expertise. \\ \hline
4 & Well-funded adversary with specialised quantum expertise or dedicated laboratory hardware. \\ \hline
5 & Nation-state-class adversary with leading-edge R\&D resources and sustained operational capacity. \\ \hline\hline
% ======================
\textbf{Exposure (\(E\))} & \textbf{Definition} \\ \hline
1 & Inaccessible or inherently resistant (e.g., physically secured and technically infeasible). \\ \hline
2 & Minor weaknesses or limited accessibility; well-mitigated or low attacker reach. \\ \hline
3 & Moderate exposure; accessible under certain conditions or with partial protections in place. \\ \hline
4 & Highly exposed; largely unprotected, feasible with available tools. \\ \hline
5 & Fully exposed and highly accessible; trivial to reach and exploit. \\ \hline
\end{tabular}
\caption{Reference scales for base Threat (adversary capability class) and Exposure (combined accessibility and vulnerability), each scored 1--5.}
\label{tab:tv_scores}
\end{table}

\noindent
To reflect contextual factors -- such as enhanced countermeasures, environmental conditions, or elevated adversary capability -- these base scores are combined multiplicatively with a strictly positive, technique‐specific multiplier \(m_i\). A global multiplier \(M\) may then be applied at the scenario level to adjust for overarching threat contexts (e.g., critical infrastructure deployment or state-sponsored actors).

The specific numerical values chosen for \(m_i\) and \(M\) are at the discretion of the risk engineers conducting the analysis. They must be justified based on threat intelligence, system characteristics, or empirical data, and used consistently across all scenarios being compared to ensure coherent and defensible results.

As a calibration guide, the following ranges are suggested:
\begin{itemize}[nosep]
  \item $[0.1,\,0.3]$ -- the condition nearly eliminates the step or scenario (e.g., the required technology does not exist, or a provably secure countermeasure is deployed);
  \item $(0.3,\,0.6]$ -- a strong but not absolute mitigation is in place (e.g., dedicated hardware defence that can, in principle, be circumvented);
  \item $(0.6,\,1.0)$ -- a moderate environmental or operational hindrance (e.g., adverse weather for free-space links);
  \item $1.0$ -- neutral baseline (no adjustment);
  \item $(1.0,\,1.5]$ -- a condition that moderately amplifies the step or scenario (e.g., combining multiple side-channel vectors);
  \item $(1.5,\,\mu_{\max}]$ -- a condition that strongly amplifies feasibility (e.g., state-level resources and motivation).
\end{itemize}
The range is intentionally asymmetric: targeted countermeasures can near-eliminate specific steps (e.g., decoy-state deployment against PNS, $m_i=0.3$), whereas aggravating factors represent incremental departures from the baseline already captured in the $T$ and $E$ scores and are therefore bounded more tightly.

Examples of both technique-level and global modifiers, with rationale for the chosen values, are shown in \autoref{tab:modifiers}.

\begin{table}[htb!]
\centering
\begin{tabular}{|p{4.5cm}|c|p{9cm}|}
\hline
\textbf{Condition} & \(\boldsymbol{m_i}\) or \(\boldsymbol{M}\) & \textbf{Rationale} \\ \hline
Extra Trojan-horse defence ($m_i$) & 0.5 & Specialised optical isolation or watchdog circuitry significantly reduces the feasibility of Trojan-horse attacks; set in the strong-mitigation range because, while effective, such defences can be partially bypassed by adaptive probing. \\ \hline
Hardened installation site (global \(M\)) & 0.6 & Physical access and remote attack surfaces are tightly controlled, reducing all scenario likelihoods; placed at the boundary of the strong-mitigation and moderate-hindrance ranges because physical security reduces but does not eliminate insider or supply-chain threats. \\ \hline
Free-space turbulence (weather, terrain) ($m_i$) & 0.7 & Environmental factors decrease the reliability of eavesdropping or interception attempts in free-space QKD; set in the moderate-hindrance range because turbulence degrades but does not prevent interception. \\ \hline
Combined side-channel exploit ($m_i$) & 1.2 & Exploiting classical and quantum side-channels together increases step effectiveness; set in the moderate-amplification range because the combined attack surface is wider, though each channel still requires independent effort. \\ \hline
State-sponsored adversary (global \(M\)) & 1.5 & Well-resourced actors can overcome mitigations and execute advanced techniques more reliably; set at the upper end of the amplification range to reflect access to dedicated laboratories, zero-day capabilities, and sustained operational campaigns. \\ \hline
Required technology not yet demonstrated ($m_i$) & 0.2 & The attack step depends on hardware capabilities (e.g., high-fidelity quantum non-demolition measurement or quantum memory at telecom wavelengths) that have not been demonstrated at the required specifications in any laboratory to date; set near the bottom of the near-elimination range because the step is currently infeasible but not ruled out by fundamental physics. \\ \hline
Decoy-state protocol deployed ($m_i$) & 0.3 & Decoy-state QKD \cite{Hwang2003,Lo_Ma_Chen_2005} renders photon-number-splitting attacks information-theoretically ineffective by enabling the receiver to detect multi-photon eavesdropping via intensity-monitoring statistics; set slightly above the technology-not-demonstrated value because effectiveness depends on correct implementation. \\ \hline
\end{tabular}
\caption{Illustrative values for technique-level (\(m_i\)) and global (\(M\)) likelihood multipliers, set by the risk analyst to reflect contextual factors.}
\label{tab:modifiers}
\end{table}

%==============================================
%==============================================
\subsection{Example: Photon-Number Splitting (PNS) Attack Kill Chain (ISO-aligned)}\label{sec:pns_example}

We illustrate the likelihood-only technique-level scoring with multiplicative modifiers, followed by scenario-level impact and 5$\times$5 matrix evaluation.

\paragraph{Step scores.}  For each step we assign base Threat \(T_i\) and Exposure \(E_i\) (\autoref{tab:tv_scores}), a technique modifier \(m_i\) (\autoref{tab:modifiers}), and compute $\ell_i$.

\begin{table}[htb!]
\centering
\small
\begin{tabular}{|c|l|l|ccc|c|c|}
\hline
$i$ & \textbf{Step (technique)}       & Phase     &$T_i$ & $E_i$ & $m_i$ & $\ell_i$ & $p_i$ \\\hline\hline
1 & Collect module info             & Knowing   & 1 & 2 & 1.0 & 2.0 & 0.080 \\
2 & Collect channel/network info    & Knowing   & 2 & 2 & 1.0 & 4.0 & 0.160 \\
3 & Develop PNS apparatus           & Knowing    & 4 & 4 & 0.2 & 3.2 & 0.128 \\
4 & Develop cyber tools             & Knowing    & 2 & 2 & 1.0 & 4.0 & 0.160 \\\hline
5 & Eavesdrop classical channel     & Entering   & 2 & 3 & 1.2 & 7.2 & 0.288 \\
6 & Tap fibre optic cable           & Entering   & 2 & 4 & 0.8 & 6.4 & 0.256 \\\hline
7 & Photon-number-splitting         & Finding    & 4 & 4 & 0.3 & 4.8 & 0.192 \\\hline
8 & Post-process quantum data       & Exploiting & 3 & 2 & 1.0 & 6.0 & 0.240 \\
9 & Abuse acquired key              & Exploiting & 3 & 2 & 1.0 & 6.0 & 0.240 \\\hline
\end{tabular}
\caption{Technique-level likelihood contributions \(\ell_i\) and step probabilities \(p_i = \ell_i / (T_{\max}\cdot E_{\max})\) for every step of the PNS kill chain.}
\label{tab:pns_steps}
\end{table}

\paragraph{Aggregate likelihood.}
Using the nine step-likelihood contributions $\{\ell_i\}$ from \autoref{tab:pns_steps}, we obtain three continuous measures $L_{\rm raw}$:

\begin{equation}
\ell = \{2.0,\;4.0,\;3.2,\;4.0,\;7.2,\;6.4,\;4.8,\;6.0,\;6.0\}.
\end{equation}

1. \textbf{Maximum:}
\begin{equation}
L_{\rm raw}^{\max}
=\max_i \ell_i
=7.2.
\end{equation}

2. \textbf{Average:}
\begin{equation}
L_{\rm raw}^{\rm avg}
=\frac{1}{9}\sum_{i=1}^{9}\ell_i
=\frac{43.6}{9}\approx4.84.
\end{equation}

3. \textbf{Probabilistic (attack-tree product):}
First convert each \(\ell_i\) into a success probability
\begin{equation}
p_i = \min\!\Bigl(1,\frac{\ell_i}{25}\Bigr)
=\{0.08,\;0.16,\;0.128,\;0.16,\;0.288,\;0.256,\;0.192,\;0.24,\;0.24\}.
\end{equation}
Then
\begin{equation}
P_{\rm chain} = \prod_{i=1}^{9} p_i \approx 2.14\times10^{-7}.
\end{equation}
With $M=1$, $P_{\rm adj}=P_{\rm chain}^{1/1}=P_{\rm chain}$. Using the logarithmic discretisation (\autoref{tab:log_bins}), $\log_{10}(P_{\rm adj})\approx -6.67$, which falls in the $(-8,\;-6]$ bin, giving $L^{\rm prod}=2$.

\noindent Here and below, the superscripts $\max$, $\rm avg$, and $\rm prod$ denote the discretised likelihood obtained via the maximum, average, and attack-tree product aggregation methods, respectively.

\paragraph{Impact and risk rating.}
In this example, the scenario-level impact is taken to be \(I=5\) (Very high). With $M=1$ and fixed bounds $L_{\min}=0$, $L_{\max}=T_{\max}\cdot E_{\max}=25$, the discretised likelihoods are:

\begin{equation}
L^{\max}=2,\quad L^{\rm avg}=1,\quad L^{\rm prod}=2.
\end{equation}
Using the 5$\times$5 matrix (\autoref{tab:risk_matrix}), the risk ratings are:
\begin{itemize}
    \item \textbf{Max-based:} \(L=2\) (Unlikely) and \(I=5\) (Very high) $\to$ cell value \(\mathbf{10}\) (Medium).
    \item \textbf{Average-based:} \(L=1\) (Very unlikely) and \(I=5\) $\to$ cell value \(\mathbf{5}\) (Medium).
    \item \textbf{Attack-tree product:} \(L=2\) (Unlikely) and \(I=5\) $\to$ cell value \(\mathbf{10}\) (Medium).
\end{itemize}

%==============================================
%==============================================
\subsection{Example: Detector-Blinding Attack Kill Chain}\label{sec:blinding_example}

To demonstrate the generalisability of the framework, we apply the same methodology to a detector-blinding attack \cite{Lydersen_2010}. In this attack, the adversary sends bright continuous-wave light into the receiver's single-photon detectors, forcing them out of single-photon (Geiger) mode into linear mode. The adversary then sends tailored ``faked-state'' pulses that deterministically control detector clicks, allowing full key extraction without introducing detectable errors.

Compared to the PNS kill chain, the detector-blinding scenario has a different resource profile: the attack apparatus uses off-the-shelf optical components rather than quantum non-demolition measurement and quantum memory.

\paragraph{Step scores.}
We assign base $T_i$, $E_i$, and $m_i$ following the same reference scales (\autoref{tab:tv_scores}, \autoref{tab:modifiers}).

\begin{table}[htb!]
\centering
\small
\begin{tabular}{|c|l|l|ccc|c|c|}
\hline
$i$ & \textbf{Step (technique)}                 & Phase     & $T_i$ & $E_i$ & $m_i$ & $\ell_i$ & $p_i$ \\\hline\hline
1 & Collect detector/module info              & Knowing   & 1 & 4 & 1.0 & 4.0 & 0.160 \\
2 & Collect channel/network info              & Knowing   & 2 & 2 & 1.0 & 4.0 & 0.160 \\
3 & Develop blinding apparatus                & Knowing   & 2 & 4 & 1.0 & 8.0 & 0.320 \\
4 & Develop cyber tools                       & Knowing   & 2 & 2 & 1.0 & 4.0 & 0.160 \\\hline
5 & Tap fibre optic cable                     & Entering  & 2 & 4 & 0.8 & 6.4 & 0.256 \\
6 & Eavesdrop classical channel               & Entering  & 2 & 3 & 1.2 & 7.2 & 0.288 \\\hline
7 & Execute blinding + faked-state injection  & Finding   & 3 & 4 & 1.2 & 14.4 & 0.576 \\\hline
8 & Post-process intercepted data             & Exploiting& 2 & 3 & 1.0 & 6.0 & 0.240 \\
9 & Abuse acquired key                        & Exploiting& 3 & 2 & 1.0 & 6.0 & 0.240 \\\hline
\end{tabular}
\caption{Technique-level likelihood contributions $\ell_i$ and step probabilities $p_i = \ell_i / (T_{\max}\cdot E_{\max})$ for the detector-blinding kill chain.}
\label{tab:blinding_steps}
\end{table}

Key differences from the PNS example: detector information (specifications, gating schemes) is often publicly available in vendor documentation ($E_1=4$); the blinding apparatus relies on standard optical components ($T_3=2$, $E_3=4$) rather than the specialised quantum hardware required for PNS; and the execution step ($T_7=3$) reflects that blinding requires advanced but not specialised quantum expertise, as demonstrated on commercial systems \cite{Lydersen_2010}. The ``Develop cyber tools'' step is shared with the PNS chain, as both attacks require tools to eavesdrop on and parse the classical channel (basis reconciliation, error correction, and privacy amplification data).

\paragraph{Aggregate likelihood.}
Using the nine step contributions from \autoref{tab:blinding_steps}:
\begin{equation}
\ell = \{4.0,\;4.0,\;8.0,\;4.0,\;6.4,\;7.2,\;14.4,\;6.0,\;6.0\}.
\end{equation}

1. \textbf{Maximum:}
\begin{equation}
L_{\rm raw}^{\max} = \max_i \ell_i = 14.4.
\end{equation}

2. \textbf{Average:}
\begin{equation}
L_{\rm raw}^{\rm avg} = \frac{1}{9}\sum_{i=1}^{9}\ell_i = \frac{60.0}{9}\approx 6.67.
\end{equation}

3. \textbf{Probabilistic (attack-tree product):}
\begin{equation}
p_i = \{0.16,\;0.16,\;0.32,\;0.16,\;0.256,\;0.288,\;0.576,\;0.24,\;0.24\},
\end{equation}
\begin{equation}
P_{\rm chain} = \prod_{i=1}^{9} p_i \approx 3.21\times10^{-6}.
\end{equation}
With $M=1$, $P_{\rm adj}=P_{\rm chain}$. Using the logarithmic discretisation (\autoref{tab:log_bins}), $\log_{10}(P_{\rm adj})\approx -5.49$, which falls in the $(-6,\;-4]$ bin, giving $L^{\rm prod}=3$.

\paragraph{Discretisation and risk rating.}
Using the same fixed bounds as the PNS example ($L_{\min}=0$, $L_{\max}=25$, with $M=1$) for max and average, the discretised likelihoods are:
\begin{equation}
L^{\max}=3,\quad L^{\rm avg}=2,\quad L^{\rm prod}=3.
\end{equation}

With scenario-level impact $I=5$ (full key extraction), the risk ratings from \autoref{tab:risk_matrix} are:
\begin{itemize}
    \item \textbf{Max-based:} $L=3$, $I=5$ $\to$ cell value $\mathbf{15}$ (High).
    \item \textbf{Average-based:} $L=2$, $I=5$ $\to$ cell value $\mathbf{10}$ (Medium).
    \item \textbf{Attack-tree product:} $L=3$, $I=5$ $\to$ cell value $\mathbf{15}$ (High).
\end{itemize}

Comparing with the PNS scenario (\autoref{sec:pns_example}), the detector-blinding attack receives higher risk ratings across all three aggregation methods, consistent with its greater practical feasibility. Under the attack-tree product, the blinding chain probability ($3.21\times10^{-6}$) is 15$\times$ higher than the PNS chain probability ($2.14\times10^{-7}$), reflecting the feasibility gap: blinding uses off-the-shelf optical components and has been demonstrated on commercial QKD systems \cite{Lydersen_2010}, while PNS requires quantum non-demolition measurement and quantum memory that do not exist at the required specifications. The modifiers capture this difference: $m=0.2$ (technology not yet demonstrated) and $m=0.3$ (decoy-state countermeasure) on the PNS chain versus $m=1.2$ (combined exploitation synergy) on the blinding execution step.

%==============================================
%==============================================
\subsection{Discussion and Practical Implications}

This technique-based model enables rapid, semi-automated risk assessment when constructing kill chains from a structured library of quantum (and classical) attack techniques. Each method is scored using consistent Threat (\(T\)) and Exposure (\(E\)) scales, with context-specific modifiers applied to reflect the real-world operating environment.

\begin{itemize}
    \item \textbf{Transparency and Repeatability:} The separation of base scores and multipliers ensures that risk assessments are both traceable and auditable. Analysts can clearly justify how each step's likelihood was derived.
    
    \item \textbf{Comparability Across Scenarios:} The fixed discretisation bounds ($L_{\min}=0$, $L_{\max}=T_{\max}\cdot E_{\max}$) produce discretised likelihood values $L$ that are directly comparable across assessments, independent of the scenario set. This supports objective prioritisation of mitigation efforts.
    
    \item \textbf{Granularity and Targeting:} Scoring each technique individually captures the wide variability between simple attacks (e.g., cable cutting) and sophisticated exploits (e.g., photon-number splitting). The most impactful or weakest step in a kill chain can be easily identified and addressed.
    
    \item \textbf{Governance and Actionability:} The matrix-based risk ratings map directly to organisational policy thresholds (e.g., \say{treat all risks \(R \geq 8\)}).

    \item \textbf{Scalability:} The scoring model is independent of the technique database size; new attack surfaces (e.g., quantum sensors, post-quantum cryptographic transitions) can be added without modifying the aggregation methodology.
\end{itemize}

\paragraph{Aggregation Method Selection.}
The divergence in risk ratings across aggregation methods is by design --- each method answers a different question and should be interpreted accordingly:
\begin{itemize}[nosep]
  \item \emph{Attack-tree product} (primary risk rating): ``What is the compound probability of full-chain success?'' It directly models the serial AND-logic of kill chains and is grounded in attack-tree and Bayesian attack graph methodology \cite{Schneier1999, NISTIR7788, Wang2007}. We recommend it as the default for all risk assessments.
  \item \emph{Maximum} (supplementary safety screening): ``Does any single step pose unacceptable risk?'' It identifies the weakest link in the chain and is appropriate for conservative, safety-critical screening. However, it is blind to chain length: a 13-step chain with one easy step scores the same as a 3-step chain with an equally easy step. It should not be interpreted as overall scenario likelihood.
  \item \emph{Average} (resource heuristic): ``What is the typical per-step difficulty?'' Because averaging compresses the dynamic range of step-level scores, this method has limited power to discriminate between scenarios and should not be used as a primary risk indicator. It is most useful for estimating the average resource investment required per attack step.
\end{itemize}
When the max-based and product-based ratings diverge --- for example, when $L^{\max}=3$ but $L^{\rm prod}=2$ --- this signals a chain with one relatively easy step among otherwise difficult ones.

\paragraph{Sensitivity to Single-Step Uncertainty.}
To illustrate the robustness of the attack-tree product, consider varying the PNS execution step by changing $m$ from 0.3 to 0.4 (a 33\% increase), yielding $\ell=4\times4\times0.4=6.4$ instead of $4.8$. Under the max method, $L^{\max}$ remains at 7.2 (unchanged, since the eavesdropping step dominates). Under the attack-tree product, $P_{\rm chain}$ changes from $2.14\times10^{-7}$ to $2.85\times10^{-7}$, remaining in the same logarithmic bin ($L^{\rm prod}=2$). The logarithmic discretisation provides natural robustness to moderate variations in individual step scores, which is desirable when expert-assigned values carry inherent imprecision.

\paragraph{Limitations and Validation.}
No real-world QKD hacking incidents have been reported to date; all known attacks on QKD, such as PNS, detector blinding, Trojan horse, laser damage, are laboratory proof-of-concept demonstrations. There is no incident database or CVE-like registry for quantum attacks, and no frequency data against which to calibrate likelihood scores. This situation is not unique to quantum risk assessment: when NIST developed SP~800-30 \cite{nist80030r1} and when the FAIR framework \cite{Jones2006FAIR} was first proposed, both relied on expert judgment for novel threats rather than historical incident data, and MITRE ATT\&CK was published as a structured methodology before extensive empirical validation \cite{mitre_attack}. ISO/IEC~27005 \cite{ISO_2022} explicitly states that likelihood may be estimated by expert judgment when statistical data is unavailable. Our $T$/$E$ scoring follows this approach, with the reference scales in \autoref{tab:tv_scores} providing anchor definitions to reduce subjectivity.

In the absence of incident-frequency data, we evaluate the methodology through two complementary forms of validation appropriate for an emerging-technology risk framework. \emph{Internal consistency:} across the two case studies, all three aggregation methods produce monotonically consistent relative rankings (detector blinding $\ge$ PNS), and the sensitivity analysis above shows that the attack-tree product rating is robust to moderate variation in individual step scores. \emph{Concurrent validity against expert literature:} the relative ranking that the framework assigns to the two scenarios matches the independent expert consensus in the published literature: detector-blinding attacks have been demonstrated end-to-end on commercial QKD systems \cite{Lydersen_2010}, whereas PNS in its original form requires high-fidelity quantum non-demolition measurement and telecom-wavelength quantum memory that have not been demonstrated at the required specifications, and is in addition rendered information-theoretically ineffective against decoy-state QKD \cite{Hwang2003,Lo_Ma_Chen_2005}. The framework recovers this consensus quantitatively (blinding chain probability $1.5\times 10^{1}$ higher than PNS under the attack-tree product) from independently assigned per-step scores, providing concurrent validity in the sense established for assessment instruments without ground-truth incident data. Formal expert elicitation (Delphi-style inter-rater study against the reference scales) and connection to emerging quantum-vulnerability databases are planned as immediate next steps.

\paragraph{Reproducibility.}
The methodology is fully specified by the formulas, scoring tables, and discretisation rules in \autoref{subsec:risk_analysis} and \autoref{subsec:tv_and_mods}. Each case study is reproducible from the inputs printed in this paper alone: the kill-chain step list (\autoref{tab:pns_steps}, \autoref{tab:blinding_steps}), the per-step Threat, Exposure, and modifier values, the global multiplier $M$, the aggregation operators of \autoref{subsec:risk_analysis}, and the discretisation bounds and bins of \autoref{tab:log_bins} and \autoref{subsec:risk_evaluation}. To enable bit-for-bit replication and re-use on new scenarios, we provide the analysis scripts that compute every numerical value reported in this paper --- step likelihoods $\ell_i$, step probabilities $p_i$, all three aggregations, the discretisation, and the final risk ratings --- as supplementary material; running them on the inputs in \autoref{tab:pns_steps} and \autoref{tab:blinding_steps} reproduces the reported $L^{\max}$, $L^{\rm avg}$, and $L^{\rm prod}$ values exactly. Re-using the methodology on a new scenario requires only a kill-chain step list and the corresponding $T_i$, $E_i$, and $m_i$ values; no scenario-dependent calibration, hidden parameters, or training data are involved.

\paragraph{Cost-Benefit Considerations.}
The quantitative risk ratings produced by this framework enable cost-benefit reasoning: organisations can compare the cost of specific countermeasures (e.g., installing optical isolators, upgrading to measurement-device-independent QKD) against the risk reduction they achieve, measured as the change in $R$. A full cost-benefit analysis is necessarily organisation-specific, depending on asset valuations, countermeasure procurement costs, and operational constraints, and is therefore beyond the scope of this paper.

\paragraph{Ethical and Legal Considerations.}
The publication of structured attack taxonomies follows the precedent of responsible disclosure established by MITRE ATT\&CK and similar threat-intelligence frameworks. The framework is designed to aid defenders; detailed attack parameters and implementation specifics are not published. Quantum technology is increasingly subject to export controls and certification requirements, as reflected in ongoing efforts by ENISA, BSI, and national security agencies to establish quantum-safe standards and certification processes \cite{position-2024, BSI_2023}.

%=========================================================
%=========================================================
%=========================================================
\section{Conclusions}\label{sec:conclusions}

As quantum communication systems move into operational networks, their security risk management must keep pace. This paper presented a framework that combines kill-chain modelling, quantitative risk assessment, and the SQOUT threat-intelligence platform.

The principal contribution is the shift from isolated vulnerability catalogues to full adversarial pathways. The framework links quantum and classical TTPs into kill chains, exposing how reconnaissance, access, and exploitation phases interact. Aggregation methods compatible with ISO/IEC 27005 translate technique-level assessments into operational risk ratings.

Contextual multipliers allow likelihood scores to reflect site hardening, environmental conditions, and adversary capabilities. SQOUT integrates the taxonomy, scoring, and analysis tools so that analysts can construct, evaluate, and update kill chains as new vulnerabilities emerge.

Several directions remain for further development. The current model assumes independence between attack steps and relies on manually assigned base scores for threat and exposure. Future work will therefore focus on introducing conditional dependencies between attack steps using probabilistic models such as Bayesian networks, as well as improving score calibration through integration with external threat-intelligence sources and expert validation. The framework can also be extended to support configurable risk-aggregation methods and to cover threats to other emerging quantum technologies, including quantum computing and quantum sensing systems.

The approach connects formal analyses of quantum attacks with established risk-assessment processes, providing a basis for operational security governance of quantum communication deployments.

%=========================================================
%=========================================================
%=========================================================
\section*{Acknowledgments}
We are grateful to Soumya Das for a thorough review of the manuscript.

%\printbibliography
\bibliography{refs} % <-- use your .bib filename without .bib
\end{document}